\documentclass[11pt]{article}
\usepackage{amssymb,amsmath,amsfonts}
\usepackage{graphicx}
\usepackage{graphics}
\usepackage{eepic,epsfig}
\usepackage{dcolumn}

\@addtoreset{equation}{section}

\makeatother

\begin{document}

\title{Vacuum currents in elliptic pseudosphere tubes}
\author{A. A. Saharian$^1$\thanks{%
Corresponding author, E-mail: saharian@ysu.am }\thinspace\ and G. V. Mirzoyan%
$^{1,2}$ \\
\textit{$^1$Institute of Physics, Yerevan State University, }\\
\textit{1 Alex Manoogian Street, 0025 Yerevan, Armenia } \vspace{0.3cm}\\
\textit{$^2$ CANDLE Synchrotron Research Institute, } \\
\textit{31 Acharyan Street, 0040 Yerevan, Armenia}}
\maketitle

\begin{abstract}
We examine the effects of spatial topology, curvature, and magnetic flux on
the vacuum expectation value (VEV) of the current density for a charged
scalar field in (2+1)-dimensional spacetime. The elliptic pseudosphere is
considered as an exactly solvable background geometry. The topological
contribution is separated in the Hadamard function for general phases in the
periodicity condition along the compact dimension. Two equivalent
expressions are provided for the component of the current density in that
direction. The corresponding VEV is a periodic function of the magnetic flux
with a period equal to the flux quantum. In the flat spacetime limit, we
recover the result for a conical space with a general value of the planar
angle deficit. Near the origin of the elliptic pseudosphere, the effect of
the spatial curvature on the vacuum current density is weak. The same
applies for small values of the length of the compact dimension. Using the
conformal relations between the elliptic pseudosphere and the
(2+1)-dimensional de Sitter spacetime with a planar angle deficit, we
determine the current densities for a conformally coupled massless scalar
field in the static and hyperbolic vacuum states of locally de Sitter
spacetime.
\end{abstract}

\bigskip

Keywords: Vacuum currents, Aharonov-Bohm effect, topological Casimir effect

\bigskip

\section{Introduction}

The dependence of physical system characteristics on dimension $D$ and the
geometry of the background space (both fundamental and effective) is an
interesting research direction. Active investigations are being conducted in
the literature on both higher ($D>3$) and lower ($D<3$) spatial dimensions.
Interest in models with $D>3$ is driven by their potential applications in
theories with extra spatial dimensions, including Kaluza-Klein, braneworld,
supergravity, and string theories.

There are two main reasons why research on low-dimensional systems is
important. First, these systems serve as simplified models of the
three-dimensional world, and the corresponding exact results can shed light
on physical processes in higher dimensions. Second, low-dimensional theories
are effective models that describe many condensed matter physics systems
(see, e.g., \cite{Torr20,Lin23}). In particular, (2+1)-dimensional field
theories well describe the long-wavelength properties of many
two-dimensional (2D) materials. A well-known example is the electronic
subsystem in Dirac materials, whose dynamics are governed by the Dirac
equation, in which the velocity of light is replaced by the Fermi velocity
of electrons \cite{Gusy07,Cast09}. A notable example of such a material is
graphene. Additional motivation comes from holographic models that connect
theories with different spatial dimensions. Studying physical effects in 2D
models clarifies the dynamics of holographically related 3D theories.

(2+1)-dimensional physical models provide a promising platform for
investigating topological and curvature-induced effects in quantum field
theory\cite{Dunn99}-\cite{Fomi18}. This research can shed light on similar
phenomena in fundamental physical theories, including those with extra
compactified spatial dimensions. The energetic band structure of 2D
materials determines the effective metric tensor and gauge potential on the
background of which low-energy quasiparticles propagate. Examples of curved
structures of 2D materials include buckyballs, fullerens, graphene nanotubes
and nanorings \cite{Dres96}-\cite{Kole06}. Spatial and temporal variations
in the microscopic characteristics of 2D crystals lead to variations in the
parameters of the band structure, resulting in effective spacetime curvature
and gauge fields. This provides an exciting opportunity to study the effects
of gravity in condensed matter systems. Lattice strain in 2D materials is an
efficient mechanism for tuning the geometric characteristics of the
background spacetime (for reviews see \cite{Vozm10}-\cite{Wei23}).
Interesting examples of analog gravity include the realization of 2D black
hole, wormhole, and cosmic string (conical) geometries, as well as the
related Hawking radiation, which have been discussed in the literature \cite%
{Iori12}-\cite{Alen21}. Analog models can also be used to study
gravitational anomalies \cite{Can16}. Another interesting effect is the
topological phase transition induced by curvature between the semimetal and
insulator phases.

In field-theoretical 2D effective models, the dependence of the properties
of the vacuum state (ground state in condensed matter systems) on background
geometry is of special interest. In quantum field theory, the vacuum is a
global concept, and its properties are sensitive to the global and local
characteristics of the bulk spacetime (see, e.g., \cite{Birr82}). In
particular, periodicity conditions on fields in models with nontrivial
spatial topology lead to Casimir-type contributions to the expectation
values of physical observables (see \cite{Most97}-\cite{Casi11} for the
boundary-induced and topological Casimir effects). In this paper, we discuss
a (2+1)-dimensional problem with nontrivial topology and spatial curvature,
where the local characteristics of the vacuum state can be evaluated
exactly. As a representative of vacuum properties, the expectation value of
the current density will be considered.

The expectation value of the current density, in addition to the
energy-momentum tensor, is an important characteristic of the vacuum state
that determines the electromagnetic backreaction of quantum effects. In the
existing literature, investigations have been conducted for various
geometries and topologies of tubes. The simplest geometry corresponds to a
flat background spacetime with toroidally compactified spatial dimensions.
The corresponding vacuum currents for scalar and Dirac fields in $D$%
-dimensional space with topology $R^{p}\times (S^{1})^{D-p}$, $p=0,1,\ldots
,D$, are studied in \cite{Bell10}-\cite{Bell15sc}. The special cases $%
(D,p)=(2,1)$ and $(D,p)=(2,0)$ correspond to cylindrical and toroidal tubes.
For the Dirac field, applications to graphene nanotubes and nanorings,
threaded by magnetic flux, are discussed. The influence of edges in
finite-length tubes is studied for Robin boundary conditions on the scalar
field and bag boundary condition on the Dirac field. The finite temperature
effects in topologies $R^{p}\times (S^{1})^{D-p}$ are discussed in \cite%
{Beze13T,Bell14T}. In this case, in addition to the current density along
the compact dimension, the nonzero charge density is generated. In the case
of helical periodicity conditions, the current density along the tube axis
appears as well \cite{Saha23}. The vacuum charge and current densities for
the Dirac field localized on planar and conical rings with circular edges
are investigated \cite{Bell16Ring,Bell20CR}. The persistent currents of a
similar nature flowing in topological insulator rings were discussed in \cite%
{Mich11}.

The conical space provides another example of flat geometries (outside the
cone apex) with nontrivial topology. In the special case $D=3$, it describes
the spacetime geometry outside an idealized straight cosmic string \cite%
{Vile94}. Graphitic cones are an example of a condensed matter realization
of conical geometry with $D=2$. The magnetic flux confined inside the cosmic
string core is a source of vacuum currents circulating in the plane
orthogonal to the string axis \cite{Srir01}-\cite{Site22}. At finite
temperatures, additional contributions to the expectation values of the
charge and current densities come from particles and antiparticles \cite%
{Moha15}-\cite{Saha25}. The influence of the background spacetime curvature
on the current density has been studied in \cite{Bell13dS}-\cite{Beze22AdS}
(see also \cite{Saha24Rev} for a review in the case of a scalar field) for
locally de Sitter (dS) and anti-de Sitter (AdS) spacetimes with toroidally
compactified spatial dimensions and in \cite{Oliv19,Beze22AdSb} for a cosmic
string on AdS spacetime. The vacuum current in Rindler spacetime partially
compactified to a torus is considered in \cite{Kota22}, assuming that the
scalar field is prepared in the Fulling-Rindler vacuum state. The
corresponding results have been used for the investigation of near-horizon
vacuum currents around cylindrical black holes. The current density on
rotationally symmetric 2D curved tubes of general geometry is investigated
in \cite{Saha24}, and an application is given for tubes having the geometry
of the Beltrami pseudosphere. Curved tubes realized by topological
insulators have been discussed recently in \cite{Graf20}-\cite{Dusa25}.

The organization of the paper is as follows. In the next section we describe
the geometry of the background spacetime and present the complete set of
mode functions for a scalar field. The expression for the Hadamard function
is derived in Section \ref{sec:Hadamard} by the summation over the scalar
modes. The vacuum expectation value (VEV) of the current density is
investigated in Section \ref{sec:Current}. The general expression is
presented and its asymptotic behavior is studied in limiting regions of the
parameters and variables. The current density for a conformally coupled
massless scalar field in the geometries conformally related to the elliptic
pseudosphere is discussed in Section \ref{sec:ConRel}. The main results are
summarized in Section \ref{sec:Conc}. In Appendix \ref{sec:App1}, we provide
an alternative representation for the Hadamard function. The conformal
relations between the elliptic pseudosphere and locally dS spacetime with a
conical defect are discussed in Appendix \ref{sec:App2}. These relations are
used in the main text to find the current density in the vacuum states of dS
spacetime corresponding to static and FLRW coordinates.

\section{Background geometry and mode functions}

\label{sec:Geometry}

\begin{figure}[tbph]
\begin{center}
\epsfig{figure=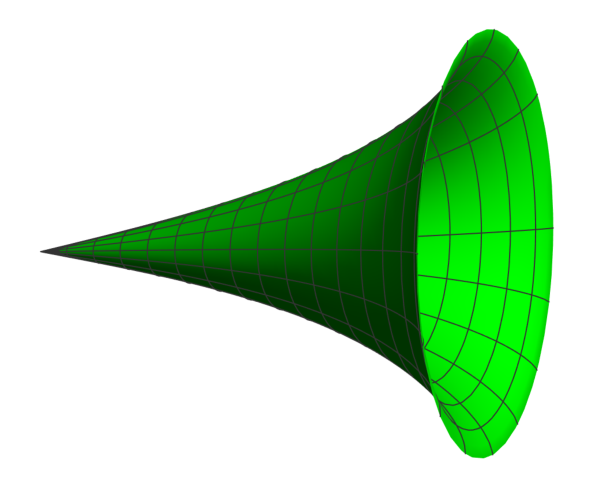,width=5cm,height=5cm}
\end{center}
\caption{Elliptic pseudosphere embedded in 3D Euclidean space.}
\label{fig1}
\end{figure}

We start the discussion by describing the background geometry and the field.
The spatial geometry under consideration corresponds to a 2-dimensional
elliptic pseudosphere with curvature radius $a$. The line element of
(2+1)-dimensional spacetime is given by%
\begin{equation}
ds^{2}=dt^{2}-a^{2}d\chi ^{2}-L^{2}\sinh ^{2}(\chi )d\phi ^{2},  \label{ds2}
\end{equation}%
where the spatial coordinates vary within the ranges $0\leq \chi <\infty $
and $0\leq \phi \leq 2\pi $. For $\chi >0$, the spatial part of the line
element describes a negative constant curvature 2D surface. For the nonzero
components of the Ricci tensor $\mathcal{R}_{ik}$ and Ricci scalar $\mathcal{%
R}$, one has%
\begin{equation}
\mathcal{R}_{1}^{1}=\mathcal{R}_{2}^{2}=-\frac{1}{a^{2}},\;\mathcal{R}=-%
\frac{2}{a^{2}}.  \label{R11}
\end{equation}%
Only a part of the elliptic pseudosphere can be embedded in a 3-dimensional
Euclidean space. This part corresponds to the range $0\leq \chi \leq \,%
\mathrm{arccosh}(a/L)$ of the coordinate $\chi $. In Fig. \ref{fig1}, the
part of the elliptic pseudosphere embedded in 3D Euclidean space is plotted
for $a/L=4$. 

We consider a complex scalar field $\varphi (x)$ in (2+1)-dimensional
spacetime with the metric tensor $g_{ik}$ determined from (\ref{ds2}) in the
coordinate system $x^{i}=(t,\chi ,\phi )$: 
\begin{equation}
g_{ik}=\mathrm{diag}(1,-a^{2},-L^{2}\sinh ^{2}\chi ).  \label{gik}
\end{equation}%
Assuming the presence of an external classical gauge field $A_{k}$, the
dynamics of the field are governed by the equation

\begin{equation}
(g^{ik}D_{i}D_{k}+m^{2}+\xi \mathcal{R})\varphi (x)=0,  \label{Feq}
\end{equation}%
with the gauge-extended covariant derivative operator $D_{k}=\nabla
_{k}+ieA_{k}$ and the curvature coupling parameter $\xi $. Here, $\nabla
_{k} $ stands for the standard covariant derivative corresponding to the
metric tensor $g_{ik}$. The special cases $\xi =0$ and $\xi =1/8$ realize
the two physically central regimes: minimal coupling ($\xi =0$ ) provides
the baseline dynamics without extra curvature interaction, while $\xi =1/8$
is the conformal value in 2+1 dimensions, ensuring conformal invariance for
a massless scalar and enabling mappings between curved and flat background
geometries. Moreover, since $\mathcal{R}=\mathrm{const}<0$ for the elliptic
pseudosphere, the effective mass term $m_{\mathrm{eff}}^{2}=m^{2}+\xi 
\mathcal{R}$ shows that varying $\xi $ between these cases shifts spectra
and hence the VEVs of physical quantities. The nontrivial spatial topology
requires a specification of a periodicity condition on the field operator
along the $\phi $-direction. Here we consider a condition with a constant
phase $\tilde{\alpha}_{p}$: 
\begin{equation}
\varphi \left( t,\chi ,\phi +2\pi \right) =e^{i\tilde{\alpha}_{p}}\varphi
\left( t,\chi ,\phi \right) .  \label{Percond}
\end{equation}%
As shown below for the example of current density, the physical
characteristics depend on the phase $\tilde{\alpha}_{p}$, being periodic
functions with the period $2\pi $.

For an external gauge field, a simple configuration will be considered with
the vector potential having the covariant components $A_{k}=(0,0,A_{2})$, $%
A_{2}=\mathrm{const}$, in the coordinate system $(t,\chi ,\phi )$. For this
configuration, the magnetic field strength is zero on the tube and the
effect of the gauge field on physical observables is of the Aharonov-Bohm
type. The gauge transformation $(\varphi ,A_{k})\rightarrow (\varphi
^{\prime },A_{k}^{\prime })$, with new fields $\varphi ^{\prime
}(x)=e^{ie\varkappa (x)}\varphi (x)$,$\;A_{k}^{\prime }=A_{k}-\partial
_{k}\varkappa (x)$, and the function $\varkappa (x)=A_{2}\phi $, leads to $%
A_{k}^{\prime }=0$. However, the vector potential $A_{2}$ does not disappear
from the problem. The gauge transformation modifies the phase in the
periodicity condition for the new field $\varphi ^{\prime }(x)$:%
\begin{equation}
\varphi ^{\prime }(t,\chi ,\phi +2\pi )=e^{i\alpha _{p}}\varphi ^{\prime
}(t,\chi ,\phi ),  \label{Per}
\end{equation}%
where the new phase is given as%
\begin{equation}
\alpha _{p}=\tilde{\alpha}_{p}+2\pi eA_{2}.  \label{alf}
\end{equation}%
The shift in the phase is interpreted in terms of the magnetic flux $\Phi
=-eA_{2}\Phi _{0}$ enclosed by the tube. Here, $\Phi _{0}=2\pi /e$ is the
flux quantum. Without loss of generality, we can discuss the problem in the
new gauge $(\varphi ^{\prime },A_{k}^{\prime }=0)$ with periodicity
condition (\ref{Per}), omitting the prime. The physical quantities will
depend on the parameters $\tilde{\alpha}_{p}$ and $A_{2}$ in the form of the
combination (\ref{alf}).

The objective of this paper is to examine the VEV of the scalar field
current density 
\begin{equation}
j_{k}(x)=ie[\varphi ^{\dagger }(x)D_{k}\varphi (x)-(D_{k}\varphi
(x))^{\dagger }\varphi (x)],  \label{jk}
\end{equation}%
in the background geometry previously outlined. In the gauge under
consideration, we have $D_{k}=\partial _{k}$. The VEVs of physical
characteristics, bilinear in the field operator, can be obtained from the
two-point functions and their derivatives in the coincidence limit of the
spacetime arguments. As a two-point function, we will choose the Hadamard
function 
\begin{equation}
G(x,x^{\prime })=\langle 0|\varphi (x)\varphi ^{\dagger }(x^{\prime
})+\varphi ^{\dagger }(x^{\prime })\varphi (x)|0\rangle ,  \label{G1a}
\end{equation}%
with $|0\rangle $ being the vacuum state. This function can be expressed in
terms of the mode-sum over a complete set of positive and negative energy
solutions $\varphi _{\sigma }^{(\pm )}(x)$ of the field equation (\ref{Feq}%
), obeying the periodicity condition along the compact dimension. Here, the
set $\sigma $ of quantum numbers specifies the modes and will be given
below. The mode-sum for the Hadamard function reads 
\begin{equation}
G(x,x^{\prime })=\sum_{\sigma }\sum_{s=+,-}\,\varphi _{\sigma
}^{(s)}(x)\varphi _{\sigma }^{(s)\ast }(x^{\prime }),  \label{G1}
\end{equation}%
where $\sum_{\sigma }$ is understood as summation over discrete quantum
numbers and integration over continuous components of $\sigma $.

In the problem at hand, the mode functions can be taken in the form%
\begin{equation}
\varphi _{\sigma }^{(\pm )}(x)=e^{ik_{n}\phi \mp i\omega t}z_{\sigma }(\chi
),\;k_{n}=n+\frac{\alpha _{p}}{2\pi },\;n=0,\pm 1,\pm 2,\ldots ,
\label{phisol}
\end{equation}%
where the eigenvalues of the momentum $k_{n}$ along the angular direction
are determined from the condition (\ref{Per}). The differential equation for
the function $z_{\sigma }(w)$ is obtained from the field equation (\ref{Feq}%
):%
\begin{equation}
\frac{\left[ \sinh (\chi )z_{\sigma }^{\prime }(\chi )\right] ^{\prime }}{%
\sinh \chi }+\left( 2\xi +a^{2}\lambda ^{2}-\frac{\mu _{n}^{2}}{\sinh
^{2}\chi }\right) z_{\sigma }(\chi )=0,  \label{Feq3}
\end{equation}%
where $\lambda ^{2}=\omega ^{2}-m^{2}$ and 
\begin{equation}
\mu _{n}=\frac{ak_{n}}{L}=\frac{a}{L}\left( n+\frac{\alpha _{p}}{2\pi }%
\right) .  \label{mun}
\end{equation}%
Introducing a new spatial coordinate $r$ in accordance with%
\begin{equation}
r=\cosh \chi ,\;1\leq r<\infty ,  \label{r}
\end{equation}%
the equation (\ref{Feq3}) is written in the form of the associated Legendre
equation 
\begin{equation}
(r^{2}-1)z_{\sigma }^{\prime \prime }(r)+2rz_{\sigma }^{\prime }(r)+\left(
2\xi +\lambda ^{2}a^{2}-\frac{\mu _{n}^{2}}{r^{2}-1}\right) z_{\sigma }(r)=0.
\label{Eqz2}
\end{equation}%
The solution of this equation, finite at $r=1$ ($\chi =0$), is expressed in
terms of the associated Legendre function of the first kind $P_{\nu }^{\mu
}(r)$ \cite{Abra72,Olve10} as%
\begin{equation}
z_{\sigma }(r)=C_{\sigma }P_{iy-\frac{1}{2}}^{-|\mu _{n}|}(r),  \label{zsigP}
\end{equation}%
where 
\begin{equation}
y=\sqrt{\lambda ^{2}a^{2}+2\xi -\frac{1}{4}}=a\sqrt{\omega ^{2}-\omega
_{m}^{2}},  \label{y}
\end{equation}%
with 
\begin{equation}
\omega _{m}^{2}=m^{2}+\frac{1}{a^{2}}\left( \frac{1}{4}-2\xi \right) .
\label{omm}
\end{equation}%
For a conformally coupled field $\omega _{m}=m$. For the modes (\ref{phisol}%
), the set of quantum numbers is specified as $\sigma =(\omega ,n)$.

Note that, introducing a new axial coordinate $R$ in accordance with 
\begin{equation}
R=\sqrt{r^{2}-1}=\sinh \chi ,\;0\leq R<\infty ,  \label{rp}
\end{equation}%
the line element (\ref{ds2}) is written in the form 
\begin{equation}
ds^{2}=dt^{2}-\frac{a^{2}dR^{2}}{1+R^{2}}-L^{2}R^{2}d\phi ^{2}.  \label{ds2b}
\end{equation}%
For $L=a$, this line element corresponds to a static (2+1)-dimensional
Friedmann-Lema\^{\i}tre-Robertson-Walker (FLRW) open universe with the scale
factor $a$. For general $L$, introducing $\phi ^{\prime }=L\phi /a$, the
line element (\ref{ds2b})\ takes the FLRW form. However, now the angular
coordinate varies in the range $0\leq \phi ^{\prime }\leq 2\pi L/a$. This
corresponds to a planar angle deficit (for $L<a$) or excess ($L>a$). Hence,
for $L\neq a$, the line element (\ref{ds2b}) can be considered as a conical
version of the static FLRW universe.

The normalization coefficient $C_{\sigma }$ in (\ref{zsigP}) is determined
from the condition 
\begin{equation}
\int_{0}^{\infty }d\chi \,\sinh \chi \int_{0}^{2\pi }d\phi \,\varphi
_{\sigma }^{(\pm )}(x)\varphi _{\sigma ^{\prime }}^{(\pm )\ast }(x)=\frac{%
\delta _{nn^{\prime }}\delta (\omega -\omega ^{\prime })}{2aL\omega }.
\label{norm}
\end{equation}%
This is reduced to the orthonormalization condition 
\begin{equation}
\int_{1}^{\infty }dr\,z_{(\omega ,n)}(r)z_{(\omega ^{\prime },n)}^{\ast }(r)=%
\frac{\delta (\omega -\omega ^{\prime })}{4\pi aL\omega },  \label{norm2}
\end{equation}%
for the function $z_{\sigma }(r)$. The integral in this condition is
evaluated by using the formula (see also \cite{Bell14Sph})%
\begin{equation}
\int_{1}^{\infty }dr\,P_{iy-\frac{1}{2}}^{-|\mu _{n}|}(r)P_{iy^{\prime }-%
\frac{1}{2}}^{-|\mu _{n}|}(r)=\frac{\pi \delta (y-y^{\prime })}{y\sinh (\pi
y)|\Gamma (iy+|\mu _{n}|+\frac{1}{2})|^{2}}.  \label{IntPP}
\end{equation}%
For the normalization coefficient we get%
\begin{equation}
\left\vert C_{\sigma }\right\vert ^{2}=\frac{\sinh (\pi y)}{4\pi ^{2}aL}%
|\Gamma (iy+|\mu _{n}|+\frac{1}{2})|^{2}.  \label{Csig}
\end{equation}%
Hence, the complete set of mode functions is expressed as 
\begin{equation}
\varphi _{\sigma }^{(\pm )}(x)=C_{\sigma }e^{ik_{n}\phi \mp i\omega t}P_{iy-%
\frac{1}{2}}^{-|\mu _{n}|}(r),  \label{Modes1}
\end{equation}%
with $k_{n}$ from (\ref{phisol}) and the normalization coefficient from (\ref%
{Csig}).

\section{Hadamard function}

\label{sec:Hadamard}

Plugging the modes (\ref{Modes1}) in the mode-sum representation (\ref{G1}),
the Hadamard function is expressed as%
\begin{equation}
G(x,x^{\prime })=\frac{1}{2\pi ^{2}L}\sum_{n=-\infty }^{+\infty
}e^{ik_{n}\Delta \phi }\int_{0}^{\infty }dy\,\frac{y\sinh (\pi y)}{\sqrt{%
y^{2}+\nu _{m}^{2}}}|\Gamma (|\mu _{n}|+\frac{1}{2}+iy)|^{2}P_{iy-\frac{1}{2}%
}^{-|\mu _{n}|}(r)P_{iy-\frac{1}{2}}^{-|\mu _{n}|}(r^{\prime })\cos (\omega
\Delta t),  \label{G12}
\end{equation}%
with $\Delta \phi =\phi -\phi ^{\prime }$, $\Delta t=t-t^{\prime }$. Here, $%
\omega =\sqrt{y^{2}/a^{2}+\omega _{m}^{2}}$ and the notation%
\begin{equation}
\nu _{m}=\sqrt{m^{2}a^{2}+1/4-2\xi },  \label{num}
\end{equation}%
is introduced. In the discussion below we assume that $\nu _{m}^{2}\geq 0$.
This condition is obeyed for the important special cases of minimally and
conformally coupled fields.

First let us consider the special case $\alpha _{p}=0$ and $L=a$. As already
mentioned above, this corresponds to static FLRW model in (2+1)-dimensional
spacetime. The expression (\ref{G12}) is transformed to the simpler form%
\begin{equation}
G(x,x^{\prime })=\frac{1}{\pi ^{2}a}\int_{0}^{\infty }dy\,\frac{y\sinh (\pi
y)}{\sqrt{y^{2}+\nu _{m}^{2}}}\cos (\omega \Delta t)\sideset{}{'}{\sum}%
_{n=0}^{\infty }\cos \left( n\Delta \phi \right) |\Gamma (n+\frac{1}{2}%
+iy)|^{2}P_{iy-\frac{1}{2}}^{-n}(r)P_{iy-\frac{1}{2}}^{-n}(r^{\prime }),
\label{Gflrw}
\end{equation}%
where the prime on the summation sign means that the term $n=0$ should be
taken with coefficient 1/2. The summation over $n$ can be done by using the
addition formula%
\begin{align}
& \overset{\infty }{\underset{n=0}{\sum }}\left( n+\frac{l}{2}\right) \Gamma
\left( \frac{l}{2}\right) C_{n}^{\frac{l}{2}}\left( \cos \Delta \phi \right)
\left\vert \Gamma \left( \frac{l+1}{2}+n+iy\right) \right\vert ^{2}P_{iy-%
\frac{1}{2}}^{-n-\frac{l}{2}}\left( r\right) P_{iy-\frac{1}{2}}^{-n-\frac{l}{%
2}}\left( r^{\prime }\right)  \notag \\
& \bigskip =\left( \frac{RR^{\prime }}{2}\right) ^{\frac{l}{2}}\left\vert
\Gamma \left( iy+\frac{l+1}{2}\right) \right\vert ^{2}\frac{P_{iy-\frac{1}{2}%
}^{-\frac{l}{2}}\left( \bar{u}\right) }{\left( \bar{u}^{2}-1\right) ^{\frac{l%
}{4}}},  \label{AdP}
\end{align}%
with $C_{n}^{l/2}\left( x\right) $ being the Gegenbauer polynomial and 
\begin{equation}
\bar{u}=rr^{\prime }-RR^{\prime }\cos \Delta \phi .  \label{u}
\end{equation}%
The summation formula (\ref{AdP}) is obtained in \cite{Saha21hyp} by using
the addition theorem for the associated Legendre functions from \cite{Henr55}%
. In the special case $l=0$, by taking into account that $C_{0}^{l/2}(\cos
\Delta \phi )=1$ and%
\begin{equation}
\lim_{l\rightarrow 0}\Gamma \left( \frac{l}{2}\right) C_{n}^{\frac{l}{2}%
}\left( \cos \Delta \phi \right) =\frac{2}{n}\cos (n\Delta \phi ),\;n\neq 0,
\label{RelGeg}
\end{equation}%
one finds%
\begin{equation}
\sideset{}{'}{\sum}_{n=0}^{\infty }\cos (n\Delta \phi )\left\vert \Gamma
\left( \frac{1}{2}+n+iy\right) \right\vert ^{2}P_{iy-\frac{1}{2}}^{-n}\left(
u\right) P_{iy-\frac{1}{2}}^{-n}\left( u^{\prime }\right) =\frac{\pi P_{iy-%
\frac{1}{2}}\left( \bar{u}\right) }{2\cosh \left( \pi y\right) }.
\label{AdP1}
\end{equation}%
With this result, for the Hadamard function in the special case under
consideration we get%
\begin{equation}
G(x,x^{\prime })=\frac{1}{\pi a}\int_{0}^{\infty }dy\,\frac{y\tanh (\pi y)}{%
\sqrt{y^{2}+\nu _{m}^{2}}}\cos (\omega \Delta t)P_{iy-1/2}\left( \bar{u}%
\right) .  \label{Gsp}
\end{equation}%
In this special case, the space is maximally symmetric and the dependence of
the two-point function on the spatial points is expressed in terms of the
geodesic distance between the points deremined by (\ref{u}).

Now we return to the general case of the parameters $\alpha _{p}$ and $L$.
For the further transformation of the Hadamard function (\ref{G12}) we apply
to the sum over $n$ in (\ref{G12}) the summation formula \cite{Bell10} 
\begin{equation}
\sum_{n=-\infty }^{+\infty }g(k_{n})f(|k_{n}|)=\int_{0}^{\infty
}du[g(u)+g(-u)]f(u)+i\int_{0}^{\infty }du[f(iu)-f(-iu)]\sum_{s=\pm 1}\frac{%
g(isu)}{e^{2\pi u+is\alpha _{p}}-1}.  \label{SumForm}
\end{equation}%
with $g(u)=e^{iu\Delta \phi }$ and%
\begin{equation}
f(u)=\Gamma \left( \frac{au}{L}+\frac{1}{2}+iy\right) \Gamma \left( \frac{au%
}{L}+\frac{1}{2}-iy\right) P_{iy-\frac{1}{2}}^{-au/L}(r)P_{iy-\frac{1}{2}%
}^{-au/L}(r^{\prime }).  \label{fu}
\end{equation}%
The Hadamard function is splitted as%
\begin{align}
G(x,x^{\prime })& =G_{0}(x,x^{\prime })+\frac{i}{2\pi ^{2}a}\int_{0}^{\infty
}dy\,\frac{y\sinh (\pi y)}{\sqrt{y^{2}+\nu _{m}^{2}}}\cos (\omega \Delta
t)\int_{0}^{\infty }dz\sum_{s=\pm 1}\frac{e^{-sLz\Delta \phi /a}}{e^{2\pi
Lz/a+is\alpha _{p}}-1}  \notag \\
& \times \sum_{j=\pm 1}j\Gamma (jiz+\frac{1}{2}+iy)\Gamma (jiz+\frac{1}{2}%
-iy)P_{iy-\frac{1}{2}}^{-jiz}(r)P_{iy-\frac{1}{2}}^{-jiz}(r^{\prime }),
\label{G2}
\end{align}%
where%
\begin{align}
G_{0}(x,x^{\prime })& =\frac{1}{\pi ^{2}a}\int_{0}^{\infty }dy\,\frac{y\sinh
(\pi y)}{\sqrt{y^{2}+\nu _{m}^{2}}}\cos (\omega \Delta t)\int_{0}^{\infty
}dz\cos \left( Lz\Delta \phi /a\right)   \notag \\
& \times |\Gamma (z+\frac{1}{2}+iy)|^{2}P_{iy-\frac{1}{2}}^{-z}(r)P_{iy-%
\frac{1}{2}}^{-z}(r^{\prime }).  \label{G0}
\end{align}%
Note that the function $G_{0}(x,x^{\prime })$ corresponds to the Hadamard
function in the geometry described by the line element (\ref{ds2}), where
the direction along the coordinate $\phi $ is decompactified, with $-\infty
<\phi <+\infty $.

For the further transformation of the second term in the right-hand side of (%
\ref{G2}), we use the relations \cite{Olve10}%
\begin{equation}
\sinh (\pi y)\Gamma \left( 1/2+iy+jiz\right) \Gamma \left( 1/2-iy+jiz\right)
P_{iy-1/2}^{-jiz}(r)=ie^{jz\pi }\left[
Q_{iy-1/2}^{jiz}(r)-Q_{-iy-1/2}^{jiz}(r)\right] ,  \label{RelPQ}
\end{equation}%
for $j=\pm 1$, where $Q_{\nu }^{\mu }(r)$ is the associated Legendre
function of the second kind. Substituting this in (\ref{G2}) one gets%
\begin{align}
G(x,x^{\prime })& =G_{0}(x,x^{\prime })-\frac{1}{2\pi ^{2}a}\int_{0}^{\infty
}dz\sum_{s=\pm 1}\frac{e^{-sLz\Delta \phi /a}}{e^{2\pi Lz/a+is\alpha _{p}}-1}
\notag \\
& \times \sum_{j,\kappa =\pm 1}j\kappa \int_{0}^{\infty }dy\,\frac{y\cos
(\omega \Delta t)}{\sqrt{y^{2}+\nu _{m}^{2}}}e^{jz\pi }Q_{\kappa
iy-1/2}^{jiz}(r)P_{iy-\frac{1}{2}}^{-jiz}(r^{\prime }).  \label{G3}
\end{align}%
As the next step, in the integral over $y$ we rotate the integration contour
in the complex $y$-plane by the angle $-\pi /2$ for the term with $\kappa =+1
$ and by the angle $\pi /2$ for the part with $\kappa =-1$. This leads to
the final expression 
\begin{align}
G(x,x^{\prime })& =G_{0}(x,x^{\prime })-\frac{2}{\pi ^{2}a}\int_{0}^{\infty
}dz\sum_{s=\pm 1}\frac{e^{-sLz\Delta \phi /a}}{e^{2\pi Lz/a+is\alpha _{p}}-1}%
\int_{\nu _{m}}^{\infty }dy\,y  \notag \\
& \times \frac{\cosh (\Delta t\sqrt{y^{2}-\nu _{m}^{2}}/a)}{\sqrt{y^{2}-\nu
_{m}^{2}}}\mathrm{Im}\left[ e^{z\pi }Q_{y-\frac{1}{2}}^{iz}(r)P_{y-\frac{1}{2%
}}^{-iz}(r^{\prime })\right] ,  \label{G4}
\end{align}%
for the Hadamard function. Here, we have used the relation%
\begin{equation}
\sum_{j=\pm 1}je^{jz\pi }Q_{y-\frac{1}{2}}^{jiz}(r)P_{y-\frac{1}{2}%
}^{-jiz}(r^{\prime })=2i\mathrm{Im}\left[ e^{z\pi }Q_{y-\frac{1}{2}%
}^{iz}(r)P_{y-\frac{1}{2}}^{-iz}(r^{\prime })\right] .  \label{RelPQ2}
\end{equation}%
An alternative representation of the Hadamard function is obtained by using
the relation (\ref{relPQ5}) from Appendix \ref{sec:App1}.

\section{Current density}

\label{sec:Current}

The VEV of the current density, $\langle 0|j_{k}(x)|0\rangle \equiv \langle
j_{k}(x)\rangle $, is expressed in terms of the Hadamard function as%
\begin{equation}
\langle j_{k}(x)\rangle =\frac{i}{2}e\lim_{x^{\prime }\rightarrow
x}(\partial _{k}-\partial _{k}^{\prime })G(x,x^{\prime }).  \label{jl1}
\end{equation}%
The part of the Hadamard function $G_{0}(x,x^{\prime })$ in (\ref{G4}),
corresponding to the uncompactified geometry, does not contribute to the
current density. By using the expression for the topological contribution in
(\ref{G4}), we see that the charge density and the current density along the 
$\chi $-direction vanish, $\langle j_{l}\rangle =0$ for $l=0,1$. For the
physical component of the current density along the compact dimension, given
by $\langle j^{\phi }\rangle =\sqrt{-g_{22}}\langle j^{2}\rangle $, we get 
\begin{equation}
\langle j^{\phi }\rangle =-\frac{2e\sin \alpha _{p}}{\pi ^{2}a^{2}R}%
\int_{0}^{\infty }\frac{zdz}{\cosh \left( 2\pi Lz/a\right) -\cos \alpha _{p}}%
\int_{\nu _{m}}^{\infty }dy\,y\,\frac{\mathrm{Im}\left[ e^{z\pi }Q_{y-\frac{1%
}{2}}^{iz}(r)P_{y-\frac{1}{2}}^{-iz}(r)\right] }{\sqrt{y^{2}-\nu _{m}^{2}}}.
\label{jphi}
\end{equation}%
In the special case $L=a$ this formula describes the current density in
static 2D FLRW open model, induced by the magnetic flux passing through the
point $\chi =0$.

An alternative representation is obtained by using the relation (\ref{relPQ5}%
):%
\begin{equation}
\langle j^{\phi }\rangle =\frac{e\sin \alpha _{p}}{\pi ^{2}a^{2}R^{2}}%
\int_{0}^{\infty }\frac{z\sinh \left( \pi z\right) dz}{\cosh \left( 2\pi
Lz/a\right) -\cos \alpha _{p}}\int_{\nu _{m}}^{\infty }dy\,y\,\frac{\left[
P_{iz-1/2}^{-y}\left( \coth \chi \right) \right] ^{2}}{\sqrt{y^{2}-\nu
_{m}^{2}}}\left\vert \Gamma \left( y+\frac{1}{2}+iz\right) \right\vert ^{2}.
\label{jphi2}
\end{equation}%
Note that the current density depends on the mass and on the curvature
coupling parameter through the combination $\nu _{m}$, given by (\ref{num}).
To simultaneously describe the dependence on $m$ and $\xi $, we will present
the numerical analysis below in terms of $\nu _{m}$. For $\nu _{m}=0$ (this
includes the case of a conformally coupled massless field) the formula (\ref%
{jphi}) is simplified to%
\begin{equation}
\langle j^{\phi }\rangle =\frac{e\sin \alpha _{p}}{\pi ^{2}a^{2}R^{2}}%
\int_{0}^{\infty }dz\int_{0}^{\infty }dy\,\frac{z\sinh \left( \pi z\right) %
\left[ P_{iz-1/2}^{-y}\left( \coth \chi \right) \right] ^{2}}{\cosh \left(
2\pi Lz/a\right) -\cos \alpha _{p}}\left\vert \Gamma \left( y+\frac{1}{2}%
+iz\right) \right\vert ^{2}.  \label{jphim0}
\end{equation}%
This expression can be used to find the current density in problems
conformally related to the problem under consideration (see below). The
physical natue of the current density (\ref{jphi2}) is similar to that for
Aharonov-Bohm flux-induced persistent currents in mesoscopic metal rings
(for experiments measuring these currents and progress in theoretical
investigations, see, e.g., \cite{Bles09}-\cite{Sark25} and the references
therein).

The parameter $\alpha _{p}$ is the shift in the phase for the field operator
under the translation along the compact dimension and, of course, the
current density is periodic with respect to that parameter with the period $%
2\pi $. By taking into account the relation (\ref{alf}), we see that the
current density is a periodic function of the magnetic flux $\Phi $ with the
period of the flux quantum $\Phi _{0}$. From (\ref{jphi2}) it follows that
the current density $\langle j^{\phi }\rangle $ is positive for $0<\alpha
_{p}<\pi $ and negative in the region $-\pi <\alpha _{p}<0$. In Fig. \ref%
{fig2}, we display the current density, as a function of the phase $\alpha
_{p}$, for the value of the radial coordinate $R=2$ and for $L/a=0.5$. The
numbers near the curves correspond to the values of the parameter $\nu _{m}$%
. 
\begin{figure}[tbph]
\begin{center}
\epsfig{figure=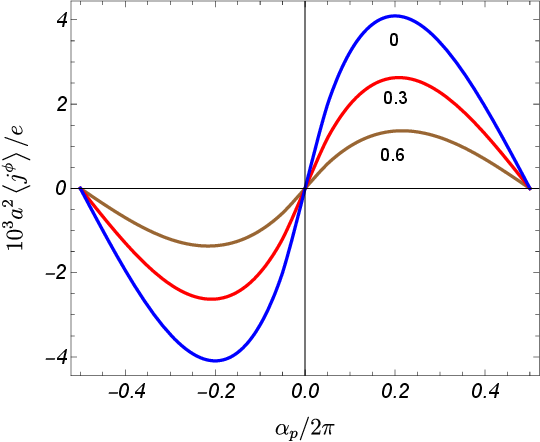,width=8cm,height=7cm}
\end{center}
\caption{The dependence of the current density on the phase $\protect\alpha %
_{p}$ for different values of $\protect\nu _{m}$ (numbers near the curves).
The graphs are plotted for $R=2$ and $L/a=0.5$. }
\label{fig2}
\end{figure}

To clarify the behavior of the current density (\ref{jphi}), as a function
of the other parameters, we consider the special and limiting cases of the
general formula.

\subsection{Flat spacetime limit}

Let us start with the flat spacetime limit. Introducing in (\ref{ds2}) a new
coordinate $w=a\chi $, we consider the limit $a\rightarrow \infty $ for
fixed $w$ and $L/a=\phi _{0}/(2\pi )$. The line element takes the form%
\begin{equation}
ds_{\mathrm{c}}^{2}=dt^{2}-dw^{2}-w^{2}d\phi ^{\prime 2},  \label{dsc2}
\end{equation}%
where $\phi ^{\prime }=L\phi /a$ and $0\leq \phi ^{\prime }\leq \phi _{0}$.
For $L=a$ this line element describes (2+1)-dimensional Minkowski spacetime.
For $L<a$ ($L>a$), (\ref{dsc2}) corresponds to a conical spacetime with
planar angle deficit (excess) $2\pi |1-L/a|$. By taking into account that $%
r=\cosh (w/a)$, we see that in the limit under consideration one has $%
r\rightarrow 1+$. In this limit and for bounded $z$ and $y$, the leading
order asymptotic 
\begin{equation*}
\mathrm{Im}\left[ e^{z\pi }Q_{y-\frac{1}{2}}^{iz}(r)P_{y-\frac{1}{2}%
}^{-iz}(r)\right] \approx \frac{1}{2z},
\end{equation*}%
is obtained by using the corresponding asymptotics for the associated
Legendre functions \cite{Olve10}. This shows that the dominant contribution
in the integral over $y$ in (\ref{jphi}) comes form the region with large $y$%
. By using the corresponding uniform asymptotic expansions for the
associated Legendre functions \cite{Olve10,Duns24}, we can see that%
\begin{equation}
e^{z\pi }Q_{y-\frac{1}{2}}^{iz}(r)P_{y-\frac{1}{2}}^{-iz}(r)\approx \frac{%
\chi }{\sinh \chi }K_{iz}\left( y\chi \right) I_{iz}\left( y\chi \right) ,
\label{relPQ3}
\end{equation}%
where $I_{\nu }(x)$ and $K_{\nu }(x)$ are the modified Bessel functions \cite%
{Abra72}. From here it follows that 
\begin{equation}
\mathrm{Im}\left[ e^{z\pi }Q_{y-\frac{1}{2}}^{iz}(r)P_{y-\frac{1}{2}%
}^{-iz}(r)\right] \approx -\frac{\chi \sinh \left( z\pi \right) }{\pi \sinh
\chi }K_{iz}^{2}\left( y\chi \right) .  \label{relPQ4}
\end{equation}

By using (\ref{relPQ4}) and noting that in the limit under consideration $%
\chi \rightarrow 0$, we get $\lim_{a\rightarrow \infty }\langle j^{\phi
}\rangle =\langle j^{\phi }\rangle _{\mathrm{cone}}$, where%
\begin{equation}
\langle j^{\phi }\rangle _{\mathrm{cone}}=\frac{2e\sin \alpha _{p}}{\pi
^{3}w^{2}}\int_{0}^{\infty }dz\int_{0}^{\infty }du\frac{z\sinh \left( z\pi
\right) K_{iz}^{2}\left( \sqrt{u^{2}+m^{2}w^{2}}\right) }{\cosh \left( \phi
_{0}z\right) -\cos \alpha _{p}},  \label{jcone}
\end{equation}%
is the current density on a cone described by the line element (\ref{dsc2}).
For a massless field the integral over $u$ is evaluated by using the formula
(see \cite{Prud2} for the general case of the order for the Macdonald
function) 
\begin{equation}
\int_{0}^{\infty }du\,K_{iz}^{2}\left( u\right) =\frac{\pi ^{2}}{4\cosh
\left( \pi z\right) },  \label{K2int}
\end{equation}%
and we obtain%
\begin{equation}
\langle j^{\phi }\rangle _{\mathrm{cone}}=\frac{e}{w^{2}}F\left( \alpha
_{p},L/a\right) ,  \label{jconem0}
\end{equation}%
where%
\begin{equation}
F\left( \alpha _{p},L/a\right) =\frac{\sin \alpha _{p}}{2\pi ^{3}}%
\int_{0}^{\infty }\frac{x\tanh (x)dx}{\cosh \left( 2xL/a\right) -\cos \alpha
_{p}},  \label{FLa}
\end{equation}%
with $L/a=\phi _{0}/(2\pi )$.

Another expression for the current density is obtained by using the integral
representation \cite{Wats66}%
\begin{equation}
K_{iz}^{2}(y)=\int_{0}^{\infty }dx\int_{0}^{\infty }\frac{dv}{v}\cos \left(
2zx\right) e^{-\frac{v}{2}-\frac{y^{2}}{v}\left( 1+\cosh (2x)\right) }.
\label{K2rep}
\end{equation}%
Plugging this in (\ref{jcone}), after integrations over $u$ and $v$, the
following formula is obtained:%
\begin{equation}
\langle j^{\phi }\rangle _{\mathrm{cone}}=\frac{e\sin \alpha _{p}}{\pi
^{2}w^{2}}\int_{0}^{\infty }\frac{z\sinh \left( z\pi \right) dz}{\cosh
\left( \phi _{0}z\right) -\cos \alpha _{p}}\int_{0}^{\infty }dx\frac{\cos
\left( 2zx\right) }{\cosh x}e^{-2wm\cosh x}.  \label{jcone2}
\end{equation}%
It can be shown that this representation is equivalent to that given in \cite%
{Saha24}. That is done by using the equality%
\begin{equation}
\int_{0}^{\infty }dx\,e^{-x-\frac{b^{2}}{2x}}\frac{K_{iz}(x)}{\sqrt{2\pi x}}%
=\int_{0}^{\infty }dx\frac{\cos \left( 2zx\right) }{\cosh x}e^{-2b\cosh x}.
\label{Rel3}
\end{equation}%
This relation is obtained by substituting the integral representation $%
K_{iz}(x)=\int_{0}^{\infty }du\,\cos \left( zu\right) e^{-x\cosh u}$ in the
left-hand side. The current density in conical spaces with general number of
spatial dimensions $D$ is investigated in \cite{Beze15}. Another
representation is obtained from the general formula in \cite{Beze15}
specifying $D=2$.

\subsection{Asymptotic and numerical analysis}

In this subsection we describe the asymptotic behavior of the current
density in limiting regions of the parameters. We then present numerical
examples. First let us consider the behavior of the current density near the
origin, $r\rightarrow 1+$ ($\chi \rightarrow 0$). As it has been mentioned
in the previous subsection, in this limit, the integral in (\ref{jphi}) is
dominated by the large values of $y$ and we can make the replacement (\ref%
{relPQ4}). Introducing a new integration variable $u=y\chi $ and assuming
that $am\chi \ll 1$, the integral over $u$ is reduced to (\ref{K2int}). To
the leading order we get%
\begin{equation}
\langle j^{\phi }\rangle \approx \frac{eF\left( \alpha _{p},L/a\right) }{%
a^{2}R^{2}},  \label{jnear0}
\end{equation}%
where $R\approx \chi \ll 1$. Note that the leading term (\ref{jnear0})
coincides with the current density (\ref{jconem0}) for a massless field on a
cone with $\phi _{0}=2\pi L/a$, where the distance from the cone apex $w$ is
replaced by the proper distance $d_{\mathrm{p}}$ for the elliptic
pseudosphere with%
\begin{equation}
d_{\mathrm{p}}=a\chi \approx aR.  \label{dp}
\end{equation}%
This shows that near the origin the effects of the spatial curvature are
week.

\begin{figure}[tbph]
\begin{center}
\begin{tabular}{cc}
\epsfig{figure=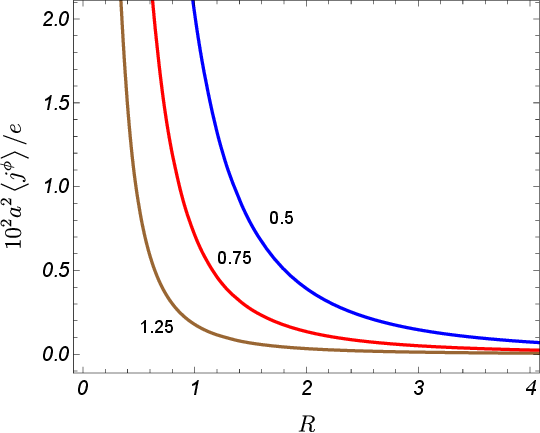,width=7.5cm,height=6cm} & \quad %
\epsfig{figure=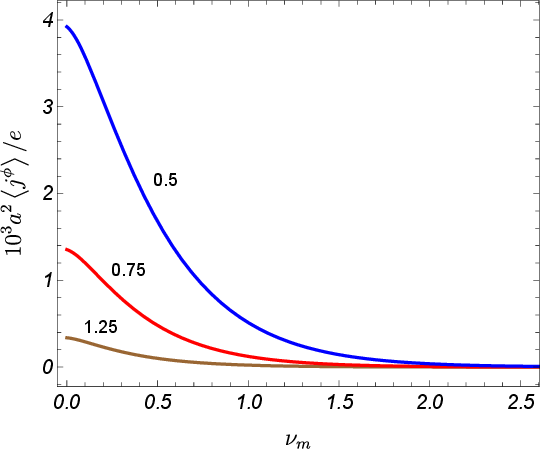,width=7.5cm,height=6cm}%
\end{tabular}%
\end{center}
\caption{The current density as a function of the coordinate $R$ for a
scalar field with $\protect\nu _{m}=0$ (left panel) and versus the parameter 
$\protect\nu _{m}$ for $R=2$ (right panel). The graphs are plotted for $%
\protect\alpha _{p}=\protect\pi /2$ and the numbers near the curves present
the values of the ratio $L/a$.}
\label{fig3}
\end{figure}

At large distances from the origin we have $r\gg 1$. In this limit, one has $%
\coth \chi \rightarrow 1+$ and it is more convenient to use the
representation (\ref{jphi2}). For the associated Legendre function we have 
\cite{Olve10} $P_{iz-1/2}^{-y}\left( \coth \chi \right) \approx e^{-y\chi
}/\Gamma \left( y+1\right) $ and the leading order term in the expression
for the current density reads%
\begin{equation}
\langle j^{\phi }\rangle \approx \frac{e\sin \alpha _{p}}{\pi ^{2}a^{2}R^{2}}%
\int_{0}^{\infty }\frac{z\sinh \left( \pi z\right) dz}{\cosh \left( 2\pi
Lz/a\right) -\cos \alpha _{p}}\int_{\nu _{m}}^{\infty }dy\,\,\frac{%
ye^{-2y\chi }}{\sqrt{y^{2}-\nu _{m}^{2}}}\frac{\left\vert \Gamma \left( y+%
\frac{1}{2}+iz\right) \right\vert ^{2}}{\Gamma ^{2}\left( y+1\right) }.
\label{jphil}
\end{equation}%
For large values of $\chi $ the dominant contribution to the integral over $y
$ comes from the region near the lower limit of integration. In the case $%
\nu _{m}=0$, we get%
\begin{equation}
\langle j^{\phi }\rangle \approx \frac{eF\left( \alpha _{p},L/a\right) }{%
a^{2}R^{2}\chi }.  \label{jphilm0}
\end{equation}%
In the limit under consideration $R\approx e^{\chi }/2$ and the current
density decays like $e^{-2d_{\mathrm{p}}/a}/d_{\mathrm{p}}$, as a function
of the proper distance. For $\nu _{m}\neq 0$, assuming that $\nu _{m}\chi
\gg 1$, the leading term is expressed as%
\begin{equation}
\langle j^{\phi }\rangle \approx \frac{2e\nu _{m}e^{-2\chi -2\nu _{m}\chi
}\sin \alpha _{p}}{\pi ^{\frac{7}{2}}a^{2}\Gamma ^{2}\left( \nu
_{m}+1\right) \sqrt{\nu _{m}\chi }}\int_{0}^{\infty }dx\,x\sinh \left(
x\right) \frac{\left\vert \Gamma \left( \nu _{m}+\frac{1}{2}+ix/\pi \right)
\right\vert ^{2}}{\cosh \left( 2Lx/a\right) -\cos \alpha _{p}},
\label{jphil2}
\end{equation}%
and the suppression, as a function of the proper distance $d_{\mathrm{p}}$,
is by the factor $e^{-2(1+\nu _{m})d_{\mathrm{p}}/a}/\sqrt{d_{\mathrm{p}}}$.
For $x\gg \pi \nu _{m}$ one has $\Gamma \left( \nu _{m}+1/2+ix/\pi \right)
\approx \pi \left( x/\pi \right) ^{2\nu _{m}}/\cosh x$ and for large $x$ the
integrand in (\ref{jphil2}) behaves as $x^{2\nu _{m}}e^{-2Lx/a}$. Note that
for both massless and massive fields, the decay of the current density is
exponential, as a function of the proper distance, at large distances from
the origin. This behavior contrasts with that for a massless field in a flat
conical space, where the decrease of the current density follows a power
law. In the left panel of Fig. \ref{fig3}, we have plotted the dependence of
the current density on the coordinate $R$ in the case $\nu _{m}=0$ for $%
\alpha _{p}=\pi /2$. The value $\nu _{m}=0$, in particular, corresponds to a
conformally coupled massless field. The dependence of the current density on
the parameter $\nu _{m}$ is depicted in the right panel of Fig. \ref{fig3}
for the same value of $\alpha _{p}$ and for $R=2$. The graphs on both panels
are plotted for the values of the ratio $L/a$ given near the corresponding
curves.

For the investigation of the asymptotic with respect to the ratio $L/a$, we
introduce a new integration variable $u=2\pi Lz/a$ in (\ref{jphi}). For $%
L/a\ll 1$, additionally assuming that $L/a\ll \nu _{m}$, it can be seen that
in the integral over $y$ the contribution from the region $y\gg 1$
dominates. In the leading order, using the asymptotic expression (\ref%
{relPQ4}) and putting $\nu _{m}=0$, the integral over $y$ is evaluated by
using (\ref{K2int}) with $z=ua/(2\pi L)$. By taking into account that $\tanh
(ua/(2\pi L))\approx 1$, for the leading order term in the current density
we find%
\begin{equation}
\langle j^{\phi }\rangle \approx \frac{e\sin \alpha _{p}}{8\pi ^{3}L^{2}R^{2}%
}\int_{0}^{\infty }\,\,\frac{u\,du}{\cosh u-\cos \alpha _{p}}.
\label{jphismallL}
\end{equation}%
Note that, in accordance with (\ref{ds2b}), $2\pi LR$ is the proper length
of the compact dimension for a given value of the coordinate $R$. The
right-hand side of (\ref{jphismallL}) coincides with the expression obtained
from (\ref{jconem0}) in the limit $L/a\ll 1$, replacing $R\rightarrow w/a$.
As seen, for small values of $L/a$ the effects of the spatial curvature are
subdominant.

\begin{figure}[tbph]
\begin{center}
\epsfig{figure=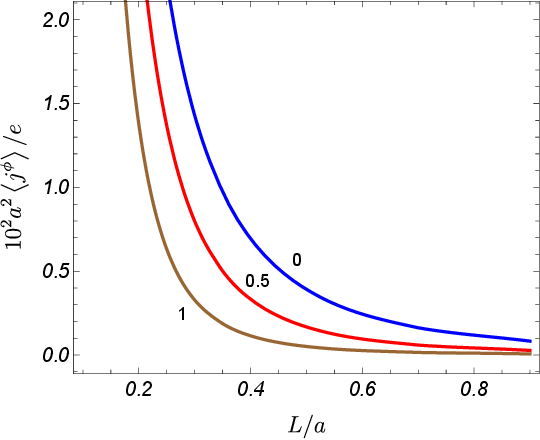,width=8cm,height=7cm}
\end{center}
\caption{The current density versus the ratio $L/a$ for different values of $%
\protect\nu _{m}$ (numbers near the curves). The graphs are plotted for $R=2$
and $\protect\alpha _{p}=\protect\pi /2$. }
\label{fig4}
\end{figure}

In the opposite limit $L/a\gg 1$, after passing to the integration over $%
u=2\pi Lz/a$, we expand the integrand over the small ratio $a/2\pi L$. To
the leading order, one obtains%
\begin{equation}
\langle j^{\phi }\rangle \approx \frac{e\sin \left( \alpha _{p}\right)
(a/L)^{3}}{8\pi ^{4}a^{2}R^{2}}\int_{0}^{\infty }\frac{u^{2}du}{\cosh u-\cos
\alpha _{p}}\int_{\nu _{m}}^{\infty }dy\,y\,\frac{\left[ P_{-1/2}^{-y}\left(
\coth \chi \right) \right] ^{2}}{\sqrt{y^{2}-\nu _{m}^{2}}}\Gamma ^{2}\left(
y+\frac{1}{2}\right) .  \label{jphilargeL}
\end{equation}%
For large $y$ one has $P_{-1/2}^{-y}\left( \coth \chi \right) \approx
e^{-y\chi }/\Gamma \left( y+1\right) $ and the integral is exponentially
convergent in the upper limit. Hence, the dimensionless quantity $%
a^{2}\langle j^{\phi }\rangle $ behaves like $(L/a)^{-2}$ for $L/a\ll 1$ and
like $\left( L/a\right) ^{-3}$ in the region $L/a\gg 1$. This quantity, as a
function of the ratio $L/a$, is plotted in Fig. \ref{fig4} for the values of
the parameter $\nu _{m}$ given near the curves. For the remaining parameters
the values $R=2$ and $\alpha _{p}=\pi /2$ are chosen. 

\section{Current density in problems conformally related to the elliptic
pseudosphere}

\label{sec:ConRel}

Consider the geometry with the line element $d\,\bar{s}^{2}$, conformally
related to the spacetime described by (\ref{ds2}) with the conformal factor $%
\Omega ^{2}(x)$, $d\,\bar{s}^{2}=\Omega ^{2}(x)ds^{2}$. In (2+1)-dimensional
spacetimes, the conformally coupled massless fields in the corresponding
problems are connected by the relation $\bar{\varphi}(x)=\varphi (x)/\Omega
^{1/2}(x)$. From here we get the relation $\langle \bar{j}_{k}(x)\rangle
=\langle j_{k}(x)\rangle /\Omega (x)$ between the covariant components. For
the physical component of the current density in problems conformally
related to the geometry under consideration, one obtains 
\begin{equation}
\langle \bar{j}^{\phi }\rangle =\Omega ^{-2}(x)\langle j^{\phi }\rangle ,
\label{jrelcc}
\end{equation}%
where $\langle j^{\phi }\rangle $ is given by (\ref{jphim0}). Here, it is
assumed that the field in the geometry with $d\,\bar{s}^{2}$ is prepared in
the vacuum state which is conformal to the vacuum in the problem with the
elliptic pseudosphere.

The conformal relation between the elliptic pseudosphere and
(2+1)-dimensional dS spacetime with angle deficit is discussed in Appendix %
\ref{sec:App2} (see also \cite{Iori14}). For the vacuum state in $\mathrm{dS}%
_{3}$ spacetime corresponding to static coordinates with the line element%
\begin{equation}
ds_{\mathrm{st}}^{2}=\left( 1-r_{\mathrm{st}}^{2}\right) dt_{\mathrm{st}%
}^{2}-a^{2}\left( \frac{dr_{\mathrm{st}}^{2}}{1-r_{\mathrm{st}}^{2}}+r_{%
\mathrm{st}}^{2}d\phi ^{\prime 2}\right) ,  \label{ds2st2}
\end{equation}%
one has $\Omega ^{2}(x)=\Omega _{\mathrm{st}}^{2}(x)=1-r_{\mathrm{st}}^{2}$
(see (\ref{ds2st})), where $r_{\mathrm{st}}=\tanh \chi _{\mathrm{st}}$. In
accordance with (\ref{jrelcc}), the corresponding current density for a
conformally coupled massless field reads 
\begin{equation}
\langle j^{\phi }\rangle _{\mathrm{st}}^{\mathrm{dS}}=\frac{e\sin \alpha _{p}%
}{\pi ^{2}a^{2}r_{\mathrm{st}}^{2}}\int_{0}^{\infty }dz\int_{0}^{\infty }dy\,%
\frac{z\sinh \left( \pi z\right) \left[ P_{iz-1/2}^{-y}\left( 1/r_{\mathrm{st%
}}\right) \right] ^{2}}{\cosh \left( 2\pi Lz/a\right) -\cos \alpha _{p}}%
\left\vert \Gamma \left( y+\frac{1}{2}+iz\right) \right\vert ^{2}.
\label{j2dS}
\end{equation}%
This formula presents the current density in the quantum state of the scalar
field corresponding to the static vacuum in dS spacetime. The asymptotics of
the expression (\ref{j2dS}) near the origin ($r_{\mathrm{st}}\ll 1$) and
near the horizon ($1-r_{\mathrm{st}}\ll 1$) are obtained from the results
given above expressed in terms of the new radial coordinate $r_{\mathrm{st}}$%
: 
\begin{align}
\langle j^{\phi }\rangle _{\mathrm{st}}^{\mathrm{dS}}& \approx \frac{%
eF\left( \alpha _{p},L/a\right) }{(ar_{\mathrm{st}})^{2}},\;r_{\mathrm{st}%
}\ll 1,  \notag \\
\langle j^{\phi }\rangle _{\mathrm{st}}^{\mathrm{dS}}& \approx -\frac{%
2eF\left( \alpha _{p},L/a\right) }{a^{2}\ln \left[ (1-r_{\mathrm{st}})/2%
\right] },\;1-r_{\mathrm{st}}\ll 1.  \label{j2dSas}
\end{align}%
The proper distance from the origin is given by $d_{\mathrm{p}}=a\arcsin (r_{%
\mathrm{st}})$, with the variation range $0\leq d_{\mathrm{p}}\leq \pi a/2$.

Another conformal relation takes place between the elliptic pseudosphere and 
$\mathrm{dS}_{3}$ with an angle deficit, described in FLRW coordinates with
the line element%
\begin{equation}
ds_{\mathrm{c}}^{2}=dt_{\mathrm{cs}}^{2}-a^{2}\sinh ^{2}\left( t_{\mathrm{cs}%
}/a\right) (d\chi _{\mathrm{c}}^{2}+\sinh ^{2}\chi _{\mathrm{c}}d\phi
^{\prime 2}).  \label{ds2cc2}
\end{equation}%
For the corresponding conformal factor one has (see (\ref{ds2cc})) $\Omega _{%
\mathrm{c}}^{2}(x)=\sinh ^{2}\left( t_{\mathrm{cs}}/a\right) $. The set of
mode functions for a conformally coupled massless field is given by $\varphi
_{\mathrm{(c)}\sigma }^{(\pm )}(x)=\varphi _{\sigma }^{(\pm )}(x)/\sqrt{%
\sinh \left( t_{\mathrm{cs}}/a\right) }$, where $\varphi _{\sigma }^{(\pm
)}(x)$ is given by (\ref{Modes1}) with $y=a\omega $. In the literature, the
vacuum state based on the quantization procedure by using the modes $\varphi
_{\mathrm{(c)}\sigma }^{(\pm )}(x)$ is known as a hyperbolic vacuum \cite%
{Saha21hyp,Pfau82,Sasa95}. The expression of the current density for a
conformally coupled scalar field in that vacuum state reads%
\begin{equation}
\langle j^{\phi }\rangle _{\mathrm{c}}^{\mathrm{dS}}=\frac{e\sin \left(
\alpha _{p}\right) \sinh ^{-2}\chi _{\mathrm{c}}}{\pi ^{2}a^{2}\sinh
^{2}\left( t_{\mathrm{cs}}/a\right) }\int_{0}^{\infty }dz\int_{0}^{\infty
}dy\,\frac{z\sinh \left( \pi z\right) \left[ P_{iz-1/2}^{-y}\left( \coth
\chi _{\mathrm{c}}\right) \right] ^{2}}{\cosh \left( 2\pi Lz/a\right) -\cos
\alpha _{p}}\left\vert \Gamma \left( y+\frac{1}{2}+iz\right) \right\vert
^{2}.  \label{j2dSc}
\end{equation}%
The near-origin and near-horizon asymptotics are give by the expressions%
\begin{align}
\langle j^{\phi }\rangle _{\mathrm{c}}^{\mathrm{dS}}& \approx \frac{eF\left(
\alpha _{p},L/a\right) }{[a\sinh \left( t_{\mathrm{cs}}/a\right) \chi _{%
\mathrm{c}}]^{2}},\;\chi _{\mathrm{c}}\ll 1,  \notag \\
\langle j^{\phi }\rangle _{\mathrm{c}}^{\mathrm{dS}}& \approx \frac{%
4eF\left( \alpha _{p},L/a\right) e^{-2\chi _{\mathrm{c}}}}{a^{2}\sinh
^{2}\left( t_{\mathrm{cs}}/a\right) \chi _{\mathrm{c}}},\;\chi _{\mathrm{c}%
}\gg 1.  \label{j2dScas}
\end{align}%
At late stages of the cosmological expansion, $t_{\mathrm{cs}}/a\gg 1$, the
current density decays like $e^{-2t_{\mathrm{cs}}/a}$. Note that both the
static and hyperbolic vacua differ from the maximally symmetric Bunch-Davies
vacuum state in dS spacetime. The vacuum currents in locally dS spacetime
with a part of spatial dimensions compactified to a torus for scalar and
Dirac fields, prepared in the Bunch-Davies state, have been studied in \cite%
{Bell13dS}. The corresponding results for 2-dimensional space were specified
in \cite{Saha24}.

\section{Conclusion}

\label{sec:Conc}

The paper studied the vacuum currents for a scalar field with general
curvature coupling parameter in (2+1)-dimension spacetime having the spatial
geometry of the elliptic pseudosphere. The metric tensor and the nonzero
components of the Ricci tensor are given by (\ref{gik}) and (\ref{R11}). The
nontrivial spatial topology requires the specification of periodicity
condition along the compact dimension. We have imposed the condition (\ref%
{Percond}) with a general constant phase. By a gauge transformation, the
phase is interpreted in terms of the magnetic flux enclosed by the compact
dimension. In the model under consideration, the properties of the vacuum
state are encoded in two-point functions. For the evaluation of the Hadamard
function we have employed the technique of the summation over a complete set
of scalar modes obeying the periodicity condition. The mode functions are
expressed in terms of the associated Legendre function of the first kind as (%
\ref{Modes1}).

The mode-sum for the Hadamard function is presented in the form (\ref{G12}).
The dependence on the mass and on the curvature coupling parameter enters
through the combination (\ref{num}). The expression for the Hadamard
function is further simplified to (\ref{Gsp}) in the special case $L=a$ and $%
\alpha _{p}=0$. This case corresponds to a (2+1)-dimensional analog of
static FLRW cosmological model with negative curvature space. In the general
case of $\alpha _{p}\neq 0$, the appearance of the nonzero vacuum current is
a topological effect and for the extraction from the two-point function of
the topological contribution we have employed the Abel-Plana type summation
formula (\ref{SumForm}). The topological part of the Hadamard function is
given by the second term in the right-hand side of (\ref{G4}). An
alternative representation is obtained by using the relation (\ref{relPQ5}).
The corresponding expression can be used for the investigation of the
topological contributions in the VEVs of physical observables bilinear in
the field operator, such as the field squared and the energy-momentum tensor.

Our main concern is the VEV of the current density, obtained from the
Hadamard function by making use of the formula (\ref{jl1}). The nonzero
component is directed along the compact dimension and the corresponding
expression is presented in two equivalent forms, Eqs. (\ref{jphi}) and (\ref%
{jphi2}). It is a periodic function of the magnetic flux enclosed by compact
dimension, with a period of the flux quantum. To clarify the behaviour of
the current density, we have considered special cases and limiting regions
of the parameters. The limit $a\rightarrow \infty $, with fixed $w=a\chi $
and $L/a$ corresponds to a conical spacetime which is flat everywhere,
except the cone apex at $w=0$ with the Dirac-delta type singularity of the
curvature tensor. The planar angle deficit or excess is given by $2\pi
|1-L/a|$. We have shown that the limiting value of the current density
obtained from the general formula coincides with the result derived
previously in the literature.

Near the origin, corresponding to $R\rightarrow 0$, the leading term in the
expansion of the current density is expressed as (\ref{jnear0}) and it
coincides with the expression for the current density in a conical space for
a massless field, where the distance from the cone apex is replaced by the
proper distance from the origin $\chi =0$ of the elliptic pseudosphere. This
shows that, near the origin, the influence of spatial curvature on the VEV
is week. At large distances, corresponding to $\chi \gg 1$, the leading
terms in the asymptotic expansion are given by (\ref{jphilm0}) and (\ref%
{jphil2}) for $\nu _{m}=0$ and $\nu _{m}\chi \gg 1$, respectively. In this
limit one has an exponential suppression by the factor $\exp [-2(1+\nu
_{m})d_{\mathrm{p}}/a]$, as a function of the proper distance $d_{\mathrm{p}%
} $. Another important dimensionless parameter in the problem is the ratio $%
L/a $. For small values of this ratio, the effect of the curvature on the
current density is weak and the leading term is given by (\ref{jphismallL}).
For $L/a\gg 1$ the asymptotic behavior is described by (\ref{jphilargeL})
and the current density decays like $\left( L/a\right) ^{-3}$ for both
massless and massive fields.

We have also described the conformal relation of the results obtained for
the elliptic pseudosphere and of the corresponding current density in
(2+1)-dimensional locally dS spacetime with a conical defect. Two types of
relations are discussed. They relate the elliptic pseudosphere to the charts
of dS$_{3}$ spacetime covered by static coordinates and the coordinates
being the (2+1)-dimensional analog of FLRW cosmological models with negative
curvatue space. The corresponding current densities for a conformally
coupled massless scalar field are given by (\ref{j2dS}) and (\ref{j2dSc})
with the near-horizon asymptotics expressed as (\ref{j2dSas}) and (\ref%
{j2dScas}). These expressions present the currents in static and hyperbolic
vacuum states for a conformally coupled massless field.

\section*{Acknowledgments}

The work was supported by the grants No. 24FP-3B021 and No. 24AA-1C013 of
the Higher Education and Science Committee of the Ministry of Education,
Science, Culture and Sport RA.

\appendix

\section{Alternative representation}

\label{sec:App1}

In this section, we will prove a relation that is used to obtain the formula
(\ref{jphi2}) for the current density. By using the Whipple's formula \cite%
{Olve10} for the associated Legendre functions, we can show that%
\begin{equation}
\mathrm{Im}\left[ e^{z\pi }Q_{y-\frac{1}{2}}^{iz}(r)P_{y-\frac{1}{2}%
}^{-iz}(r^{\prime })\right] =\frac{P_{iz-1/2}^{-y}\left( u_{\chi }\right) }{%
\sqrt{RR^{\prime }}}\mathrm{Im}\,\left[ \Gamma \left( y+\frac{1}{2}%
+iz\right) \mathbf{Q}_{iz-1/2}^{-y}\left( u_{\chi ^{\prime }}\right) \right]
,  \label{relPQ}
\end{equation}%
where $u_{\chi }=r/R=\coth \chi $ and $\mathbf{Q}_{\nu }^{\mu }(x)=e^{-\mu
\pi i}Q_{\nu }^{\mu }(x)/\Gamma (\mu +\nu +1)$. Note that the function $%
P_{iz-1/2}^{-y}\left( u_{\chi }\right) $ is real for real $y$ and $z$. As
the next step, we employ the relation \cite{Olve10} 
\begin{equation}
\mathbf{Q}_{iz-1/2}^{-y}(x)=\frac{\pi }{2\sin \left( y\pi \right) }\left[ 
\frac{P_{iz-1/2}^{y}(x)}{\Gamma \left( iz+1/2+y\right) }-\frac{%
P_{iz-1/2}^{-y}(x)}{\Gamma \left( iz+1/2-y\right) }\right]   \label{QPrel}
\end{equation}%
between the associated Legendre functions. This gives%
\begin{equation}
\mathrm{Im}\left[ e^{z\pi }Q_{y-\frac{1}{2}}^{iz}(r)P_{y-\frac{1}{2}%
}^{-iz}(r^{\prime })\right] =-\frac{\pi P_{iz-1/2}^{-y}\left( u_{\chi
}\right) P_{iz-1/2}^{-y}(u_{\chi ^{\prime }})}{2\sin \left( y\pi \right) 
\sqrt{RR^{\prime }}}\mathrm{Im}\left[ \frac{\Gamma \left( y+\frac{1}{2}%
+iz\right) }{\Gamma \left( iz+1/2-y\right) }\right] .  \label{relPQ2}
\end{equation}%
By using%
\begin{equation}
\Gamma \left( iz+1/2-y\right) =\frac{\pi }{\cos [\pi \left( y-iz\right)
]\Gamma (y+1/2-iz)},  \label{Gam1}
\end{equation}%
and taking the imaginary part for $\cos [\pi \left( y-iz\right) ]$, one finds%
\begin{equation}
\mathrm{Im}\left[ e^{z\pi }Q_{y-\frac{1}{2}}^{iz}(r)P_{y-\frac{1}{2}%
}^{-iz}(r^{\prime })\right] =-\frac{\sinh \pi z}{2\sqrt{RR^{\prime }}}%
\left\vert \Gamma \left( y+\frac{1}{2}+iz\right) \right\vert
^{2}P_{iz-1/2}^{-y}\left( \coth \chi \right) P_{iz-1/2}^{-y}\left( \coth
\chi ^{\prime }\right) .  \label{relPQ5}
\end{equation}%
Note that the functions $P_{iz-1/2}^{-y}\left( x\right) $ and $%
Q_{iz-1/2}^{-y}\left( x\right) $, with real $z$, are also known in the
literature as conical or Mehler functions \cite{Olve10}.

\section{Conformal relations between the elliptic pseudosphere and $\mathrm{%
dS}_{3}$}

\label{sec:App2}

We denote by $X^{M}$, $M=0,1,2,3$, the coordinates in (3+1)-dimensional
Minkowski spacetime with the line element $ds_{4}^{2}=\eta _{MN}dX^{M}dX^{N}$
and the metric tensor $\eta _{MN}=\mathrm{diag}(1,-1,-1,-1)$. The
(2+1)-dimensional de Sitter spacetime with curvature radius $a$, $\mathrm{dS}%
_{3}$, is defined as a hyperboloid $\eta _{MN}X^{M}X^{N}=-a^{2}$. ~The
global coordinates $(t_{\mathrm{gs}},\chi _{\mathrm{g}},\phi ^{\prime })$
that cover the entire spacetime are introduced in accordance with%
\begin{align}
& X^{0}=a\sinh (t_{\mathrm{gs}}/a),\;X^{1}=a\cosh (t_{\mathrm{gs}}/a)\cos
\chi _{\mathrm{g}},  \notag \\
& \left( X^{2},X^{3}\right) =a\cosh (t_{\mathrm{gs}}/a)\sin \chi _{\mathrm{g}%
}\,\left( \cos \phi ^{\prime },\sin \phi ^{\prime }\right) .  \label{Global}
\end{align}%
The corresponding line element reads 
\begin{equation}
ds^{2}=dt_{\mathrm{gs}}^{2}-a^{2}\cosh ^{2}(t_{\mathrm{gs}}/\alpha )\left(
d\chi _{\mathrm{g}}^{2}+\sin ^{2}\chi _{\mathrm{g}}d\phi ^{\prime 2}\right) .
\label{dsg}
\end{equation}%
In $\mathrm{dS}_{3}$, for the variation ranges of the coordinates one has $%
-\infty <t_{\mathrm{gs}}<+\infty $, $0<\chi _{\mathrm{g}}<\pi $, $0\leqslant
\phi ^{\prime }\leqslant 2\pi $. Here we will consider the geometry with an
angle deficit, assuming that $0\leq \phi ^{\prime }\leq \phi _{0}$. For $%
\chi _{\mathrm{g}}\neq 0,\pi $ the local geometry is the same as that for $%
\mathrm{dS}_{3}$ and the curvature tensor has Dirac delta function type
singularities at $\chi _{\mathrm{g}}=0,\pi $. The latter are the analog of
the singularity at the cone apex in (2+1)-dimensional conical spacetime (see
(\ref{dsc2})). With the new time coordinate $t_{\mathrm{g}}$, $0<t_{\mathrm{g%
}}/a<\pi $, defined by the relation $\sin (t_{\mathrm{g}}/a)=1/\cosh (t_{%
\mathrm{gs}}/a)$, we get a conformally static representation of the line
element:%
\begin{equation}
ds^{2}=\frac{dt_{\mathrm{g}}^{2}-a^{2}\left( d\chi _{\mathrm{g}}^{2}+\sin
^{2}\chi _{\mathrm{g}}d\phi ^{\prime 2}\right) }{\sin ^{2}(t_{\mathrm{g}}/a)}%
.  \label{dsgc}
\end{equation}

To see the conformal relation with the elliptic pseudosphere, we need a
negative curvature spatial foliation of $\mathrm{dS}_{3}$. This is realized
introducing the coordinates $(t_{\mathrm{st}},\chi _{\mathrm{st}},\phi )$, $%
-\infty <t_{\mathrm{st}}<+\infty $, $0<\chi _{\mathrm{st}}<\infty $, $0\leq
\phi \leq \pi $, in accordance with 
\begin{align}
& X^{0}=\frac{a\sinh (t_{\mathrm{st}}/a)}{\sinh \chi _{\mathrm{st}}},\;X^{1}=%
\frac{a\cosh (t_{\mathrm{st}}/a)}{\sinh \chi _{\mathrm{st}}},  \notag \\
& \left( X^{2},X^{3}\right) =a\tanh \chi _{\mathrm{st}}\,\left( \cos \phi
^{\prime },\sin \phi ^{\prime }\right) .  \label{Xst}
\end{align}%
The line element is reduced to 
\begin{equation}
ds_{\mathrm{st}}^{2}=\frac{dt_{\mathrm{st}}^{2}-a^{2}\left( d\chi _{\mathrm{%
st}}^{2}+\sinh ^{2}\chi _{\mathrm{st}}d\phi ^{2}\right) }{\cosh ^{2}\chi _{%
\mathrm{st}}}=\frac{ds^{2}(t_{\mathrm{st}},\chi _{\mathrm{st}})}{\cosh
^{2}\chi _{\mathrm{st}}}.  \label{ds2st}
\end{equation}%
Taking a new radial coordinate $r_{\mathrm{st}}=\tanh \chi _{\mathrm{st}}$,
the line element (\ref{ds2st}) is written in the form (\ref{ds2st2}), which
is the standard form of the dS line element in static coordinates (for the
conformal relation between the elliptic pseudosphere and the static chart of
dS spacetime, see also \cite{Iori14}).

Another set of coordinates, $(t_{\mathrm{c}},\chi _{\mathrm{c}},\phi )$,
with $-\infty <t_{\mathrm{c}}<0$, $0<\chi _{\mathrm{c}}<\infty $ (here the
index c stands for cosmological), covering a part of dS spacetime,
corresponds to 
\begin{align}
& X^{0}=-\frac{a\cosh \chi _{\mathrm{c}}}{\sinh \left( t_{\mathrm{c}%
}/a\right) },\;X^{1}=-a\coth \left( t_{\mathrm{c}}/a\right) ,  \notag \\
& \left( X^{2},X^{3}\right) =-\frac{a\sinh \chi _{\mathrm{c}}}{\sinh \left(
t_{\mathrm{c}}/a\right) }\,\left( \cos \phi ^{\prime },\sin \phi ^{\prime
}\right) .  \label{Xc}
\end{align}%
In these coordinates, the (2+1)-dimensional line element is expressed as%
\begin{equation}
ds_{\mathrm{c}}^{2}=\frac{dt_{\mathrm{c}}^{2}-a^{2}\left( d\chi _{\mathrm{c}%
}^{2}+\sinh ^{2}\chi _{\mathrm{c}}d\phi ^{\prime 2}\right) }{\sinh
^{2}\left( t_{\mathrm{c}}/a\right) }=\frac{ds^{2}(t_{\mathrm{c}},\chi _{%
\mathrm{c}})}{\sinh ^{2}\left( t_{\mathrm{c}}/a\right) }.  \label{ds2cc}
\end{equation}%
Introducing the synchronous time coordinate $t_{\mathrm{cs}}$, $0<t_{\mathrm{%
cs}}<\infty $, in accordance with $e^{t_{\mathrm{c}}/a}=\tanh \left( t_{%
\mathrm{cs}}/2a\right) $, for the line element (\ref{ds2cc}) we obtain the
presentation (\ref{ds2cc2}). The latter is the (2+1)-dimensional analog of
dS spacetime used in FLRW cosmological models. Note that we have the
relation $\sinh \left( t_{\mathrm{c}}/a\right) =-1/\sinh \left( t_{\mathrm{cs%
}}/a\right) $. The sets of the coordinates $(t_{\mathrm{c}},\chi _{\mathrm{c}%
},\phi )$ and $(t_{\mathrm{st}},\chi _{\mathrm{st}},\phi )$, realizing two
conformal relations, are connected by the transformation 
\begin{equation}
\tanh (t_{\mathrm{st}}/a)=\frac{\cosh \chi _{\mathrm{c}}}{\cosh \left( t_{%
\mathrm{c}}/a\right) },\;\tanh \chi _{\mathrm{st}}=-\frac{\sinh \chi _{%
\mathrm{c}}}{\sinh \left( t_{\mathrm{c}}/a\right) },  \label{RelCoord}
\end{equation}%
with the same angular coordinates $\phi ^{\prime }$.


\begin{thebibliography}{99}
\bibitem{Torr20} L.E.F.F. Torres, S. Roche, J.-C. Charlier, \textit{%
Introduction to Graphene-Based Nanomaterials} (Cambridge University Press,
Cambridge, 2020).

\bibitem{Lin23} Y.-C. Lin, et. al., Recent advances in 2D material theory,
synthesis, properties, and applications, ACS Nano \textbf{17}, 9694 (2023).

\bibitem{Gusy07} V.P. Gusynin, S.G. Sharapov, J.P. Carbotte, AC conductivity
of graphene: From tight-binding model to 2+1-dimensional quantum
electrodynamics, Int. J. Mod. Phys. B \textbf{21}, 4611 (2007).

\bibitem{Cast09} A.H. Castro Neto, F. Guinea, N.M.R. Peres, K.S. Novoselov,
A.K. Geim, The electronic properties of graphene, Rev. Mod. Phys. \textbf{81}%
, 109 (2009).

\bibitem{Dunn99} G.V. Dunne, \textit{Topological Aspects of Low Dimensional
Systems} (Springer, Berlin, 1999).

\bibitem{Imry08} Y. Imry, \textit{Introduction to Mesoscopic Physics}
(Oxford University Press, New York, USA, 2008).

\bibitem{Fomi18} V.M. Fomin (Ed.), \textit{Physics of Quantum Rings}
(Springer International Publishing, Cham, Switzerland, 2018).

\bibitem{Dres96} M.S. Dresselhaus, G. Dresselhaus, P.C. Eklund, \textit{%
Science of Fullerenes and Carbon Nanotubes} (Academic, 1996).

\bibitem{Liu02} L. Liu, G.Y. Guo, C.S. Jayanthi, S.Y. Wu, Colossal
paramagnetic moments in metallic carbon nanotori, Phys. Rev. Lett. \textbf{88%
}, 217206 (2002).

\bibitem{Shyu04} F.L. Shyu, C.C. Tsai, C.P. Chang, R.B. Chen, M.F. Lin,
Magnetoelectronic states of carbon toroids, Carbon \textbf{42}, 2879 (2004).

\bibitem{Pozr08} C. Pozrikidis, Structure of carbon nanorings, Computational
Materials Science \textbf{43}, 943 (2008).

\bibitem{Gonz93} J. Gonz\'{a}lez, F. Guinea, M.A.H. Vozmediano, The
electronic spectrum of fullerenes from the Dirac equation, Nucl. Phys. B 
\textbf{406}, 771 (1993).

\bibitem{Pudl06} M. Pudlak, R. Pincak, V.A. Osipov, Low-energy electronic
states in spheroidal fullerenes, Phys. Rev. B \textbf{74}, 235435 (2006).

\bibitem{Kole06} D.V. Kolesnikov, V.A. Osipov, The continuum gauge
field-theory model for low-energy electronic states of icosahedral
fullerenes, Eur. Phys. J. B \textbf{49}, 465 (2006).

\bibitem{Vozm10} M.A.H. Vozmediano, M.I. Katsnelson, F. Guinea, Gauge fields
in graphene, Phys. Rept. \textbf{496}, 109 (2010).

\bibitem{Naum17} G.G. Naumis, S. Barraza-Lopez, M. Oliva-Leyva, H. Terrones,
Electronic and optical properties of strained graphene and other strained 2D
materials: a review, Rep. Prog. Phys. \textbf{80}, 096501 (2017).

\bibitem{Peng20} Z. Peng, X. Chen, Y. Fan, D.J. Srolovitz, D. Lei, Strain
engineering of 2D semiconductors and graphene: from strain fields to
band-structure tuning and photonic applications, Light: Science \&
Applications \textbf{9}, 190 (2020).

\bibitem{Yang21} S.X. Yang, Y.J. Chen, C.B. Jiang, Strain engineering of
two-dimensional materials: Methods, properties, and applications, InfoMat 
\textbf{3}, 397 (2021).

\bibitem{Blun21} E. Blundo, E. Cappelluti, M. Felici, G. Pettinari, A.
Polimeni, Strain-tuning of the electronic, optical, and vibrational
properties of two-dimensional crystals, Appl. Phys. Rev. \textbf{8}, 021318
(2021).

\bibitem{Wei23} N. Wei, Y. Ding, J. Zhang, L. Li, M. Zeng, L. Fu, Curvature
geometry in 2D materials, Natl Sci Rev. \textbf{10}, 145 (2023).

\bibitem{Iori12} A. Iorio, G. Lambiase, The Hawking-Unruh phenomenon on
graphene, Phys. Lett. B \textbf{716}, 334 (2012).

\bibitem{Cvet12} M. Cvetic, G.W. Gibbons, Graphene and the Zermelo optical
metric of the BTZ black hole, Ann. Phys. (N.Y.) \textbf{327}, 2617 (2012).

\bibitem{Kand20} B.S. Kandemir, Hairy BTZ black hole and its analogue model
in graphene, Ann. Phys. (N.Y.) \textbf{413}, 168064 (2020).

\bibitem{Iori21} A. Iorio, Carbon pseudospheres and the BTZ black hole,
PoS(CORFU2021)240.

\bibitem{Iori14} A. Iorio, G. Lambiase, Quantum field theory in curved
graphene spacetimes, Lobachevsky geometry, Weyl symmetry, Hawking effect,
and all that, Phys. Rev. D \textbf{90}, 025006 (2014).

\bibitem{Naes09} S.N. Naess, A. Elgsaeter, G. Helgesen, K.D. Knudsen, Carbon
nanocones: wall structure and morphology, Sci. Tech. Adv. Mater. \textbf{10}%
, 065002 (2009).

\bibitem{Capo18} S. Capozziello, R. Pincak, E.N. Saridakis, Constructing
superconductors by graphene Chern-Simons wormholes, Ann. Phys. (N.Y.) 
\textbf{390}, 303 (2018).

\bibitem{Rojj19} T. Rojjanason, P. Burikham, K. Pimsamarn, Charged fermion
in (1 + 2)-dimensional wormhole with axial magnetic field, Eur. Phys. J. C 
\textbf{79}, 660 (2019).

\bibitem{Alen21} G. Alencar, V.B. Bezerra, C.R. Muniz, Casimir wormholes in
2+1 dimensions with applications to the graphene, Eur. Phys. J. C \textbf{81}%
, 924 (2021).

\bibitem{Can16} T. Can, Y.H. Chiu, M. Laskin, P. Wiegmann, Emergent
conformal symmetry and geometric transport properties of quantum Hall states
on singular surfaces, Phys. Rev. Lett. \textbf{117}, 266803 (2016).

\bibitem{Birr82} N.D. Birrell, P.C.W. Davies, \textit{Quantum Fields in
Curved Space} (Cambridge University Press, Cambridge, England, 1982).

\bibitem{Most97} V.M. Mostepanenko, N.N. Trunov, \textit{The Casimir Effect
and Its Applications} (Clarendon, Oxford, 1997).

\bibitem{Milt02} K.A. Milton, \textit{The Casimir Effect: Physical
Manifestation of Zero-Point Energy} (World Scientific, Singapore, 2002).

\bibitem{Bord09} M. Bordag, G.L. Klimchitskaya, U. Mohideen, V.M.
Mostepanenko, \textit{Advances in the Casimir Effect} (Oxford University
Press, New York, 2009).

\bibitem{Casi11} \textit{Casimir Physics}, edited by D. Dalvit, P. Milonni,
D. Roberts, F. da Rosa, Lecture Notes in Physics Vol. 834 (Springer-Verlag,
Berlin, 2011).

\bibitem{Bell10} S. Bellucci, A.A. Saharian, V.M. Bardeghyan, Induced
fermionic current in toroidally compactified spacetimes with applications to
cylindrical and toroidal nanotubes, Phys. Rev. D \textbf{82}, 065011 (2010).

\bibitem{Bell13} S. Bellucci, A. A. Saharian, Fermionic current from
topology and boundaries with applications to higher-dimensional models and
nanophysics, Phys. Rev. D \textbf{87}, 025005 (2013).

\bibitem{Bell15sc} S. Bellucci, A.A. Saharian, N.A. Saharyan, Casimir effect
for scalar current densities in topologically nontrivial spaces, Eur. Phys.
J. C \textbf{75}, 378 (2015).

\bibitem{Beze13T} E.R. Bezerra de Mello, A.A. Saharian, Finite temperature
current densities and Bose-Einstein condensation in topologically nontrivial
spaces, Phys. Rev. D \textbf{87}, 045015 (2013).

\bibitem{Bell14T} S. Bellucci, E.R. Bezerra de Mello, A.A. Saharian, Finite
temperature fermionic condensate and currents in topologically nontrivial
spaces, Phys. Rev. D \textbf{89}, 085002 (2014).

\bibitem{Saha23} A.A. Saharian, D.H. Simonyan, H.H. Mikayelyan, A.A.
Vantsyan, Helical vacuum currents for a scalar field in models with
nontrivial spatial topology, J. Contemp. Phys. \textbf{58}, 341 (2023).

\bibitem{Bell16Ring} S. Bellucci, A.A. Saharian, A.Kh. Grigoryan, Induced
fermionic charge and current densities in two-dimensional rings, Phys. Rev.
D \textbf{94}, 105007 (2016).

\bibitem{Bell20CR} S. Bellucci, I. Brevik, A.A. Saharian, H.G. Sargsyan, The
Casimir effect for fermionic currents in conical rings with applications to
graphene ribbons, Eur. Phys. J. C \textbf{80}, 281 (2020).

\bibitem{Mich11} P. Michetti, P. Recher, Bound states and persistent
currents in topological insulator rings, Phys. Rev. B \textbf{83}, 125420
(2011).

\bibitem{Vile94} A. Vilenkin, E.P.S. Shellard, \textit{Cosmic Strings and
Other Topological Defects} (Cambridge University Press, Cambridge, England,
1994).

\bibitem{Srir01} L. Sriramkumar, Fluctuations in the current and energy
densities around a magnetic-flux-carrying cosmic string, Classical Quantum
Gravity \textbf{18}, 1015 (2001).

\bibitem{Site09} Yu.A. Sitenko, N.D. Vlasii, Induced vacuum current and
magnetic field in the background of a cosmic string, Classical Quantum
Gravity \textbf{26}, 195009 (2009).

\bibitem{Beze10cs} E.R. Bezerra de Mello, Induced fermionic current
densities by magnetic flux in higher dimensional cosmic string spacetime,
Classical Quantum Gravity \textbf{27}, 095017 (2010).

\bibitem{Beze15} E.R. Bezerra de Mello, V.B. Bezerra, A.A. Saharian, H.H.
Harutyunyan, Vacuum currents induced by a magnetic flux around a cosmic
string with finite core, Phys. Rev. D \textbf{91}, 064034 (2015).

\bibitem{Site18} Y.A. Sitenko, V.M. Gorkavenko, Non-Euclidean geometry,
nontrivial topology and quantum vacuum effects, Universe \textbf{4}, 23
(2018).

\bibitem{Site22} Y.A. Sitenko, V.M. Gorkavenko, M.S. Tsarenkova, Magnetic
flux in the vacuum of quantum bosonic matter in the cosmic string
background, Phys. Rev. D \textbf{106}, 105010 (2022).

\bibitem{Moha15} A. Mohammadi, E.R. Bezerra de Mello, A.A. Saharian, Finite
temperature fermionic charge and current densities induced by a cosmic
string with magnetic flux, J. Phys. A: Math. Theor. \textbf{48}, No.18,
185401 (2015).

\bibitem{Bell16FT} S. Bellucci, E.R. Bezerra de Mello, E. Bragan\c{c}a, A.A.
Saharian, Finite temperature fermion condensate, charge and current
densities in a (2+1)-dimensional conical space, Eur. Phys. J. C \textbf{76},
350 (2016).

\bibitem{Saha25} A.A. Saharian, V.F. Manukyan, T.A. Petrosyan, Finite
temperature fermionic charge and current densities in conical space with a
circular edge, Phys. Rev. D \textbf{111}, 065006 (2025).

\bibitem{Bell13dS} S. Bellucci, A.A. Saharian, H.A. Nersisyan, Scalar and
fermionic vacuum currents in de Sitter spacetime with compact dimensions,
Phys. Rev. D \textbf{88}, 024028 (2013).

\bibitem{Beze15AdS} E.R. Bezerra de Mello, A.A. Saharian, V. Vardanyan,
Induced vacuum currents in anti-de Sitter space with toral dimensions, Phys.
Lett. B \textbf{741}, 155 (2015).

\bibitem{Bell17AdS} S. Bellucci, A.A. Saharian, V. Vardanyan, Fermionic
currents in AdS spacetime with compact dimensions, Phys. Rev. D \textbf{96},
065025 (2017).

\bibitem{Beze22AdS} E.R. Bezerra de Mello, W. Oliveira dos Santos, A.A.
Saharian, Finite temperature charge and current densities around a cosmic
string in AdS spacetime with compact dimension, Phys. Rev. D \textbf{106},
125009 (2022).

\bibitem{Saha24Rev} A.A. Saharian, Vacuum currents for a scalar field in
models with compact dimensions, Symmetry \textbf{16(1)}, 92 (2024).

\bibitem{Oliv19} W. Oliveira dos Santos, H.\thinspace F. Mota, E.\thinspace
R. Bezerra de Mello, Induced current in high-dimensional AdS spacetime in
the presence of a cosmic string and a compactified extra dimension, Phys.
Rev. D \textbf{99}, 045005 (2019).

\bibitem{Beze22AdSb} E.R. Bezerra de Mello, W. Oliveira dos Santos, A.A.
Saharian, Finite temperature charge and current densities around a cosmic
string in AdS spacetime with compact dimension, Phys. Rev. D \textbf{106},
125009 (2022).

\bibitem{Kota22} V.Kh. Kotanjyan, A.A. Saharian, M.R. Setare, Vacuum
currents in partially compactified Rindler spacetime with an application to
cylindrical black holes, Nucl. Phys. B \textbf{980}, 115838 (2022).

\bibitem{Saha24} A.A. Saharian, Vacuum currents in curved tubes, Phys. Rev.
D \textbf{110}, 065020 (2024).

\bibitem{Graf20} A. Graf, R. Kozlovsky, K. Richter, C. Gorini, Theory of
magnetotransport in shaped topological insulator nanowires, Phys. Rev. B 
\textbf{102}, 165105 (2020).

\bibitem{Fust24} M. F\"{u}st, D. Kochan, I.-G. Dusa, C. Gorini, K. Richter,
Dirac Landau levels for surfaces with constant negative curvature, Phys.
Rev. B \textbf{109}, 195433 (2024).

\bibitem{Dusa25} I. Dusa, D. Kochan, M. F\"{u}st, C. Gorini, K. Richter,
Hearing the shape of a Dirac drum: Dual quantum Hall states on curved
surfaces, arXiv:2503.17166.

\bibitem{Abra72} \textit{Handbook of Mathematical Functions}, edited by M.
Abramowitz and I. A. Stegun (Dover, New York, 1972).

\bibitem{Olve10} F.W. Olver et al., \textit{NIST Handbook of Mathematical
Functions} (Cambridge University Press, USA, 2010).

\bibitem{Bell14Sph} S. Bellucci, A.A. Saharian, N.A. Saharyan, Wightman
function and the Casimir effect for a Robin sphere in a constant curvature
space, Eur. Phys. J. C \textbf{74}, 3047 (2014).

\bibitem{Saha21hyp} A.A. Saharian, T.A. Petrosyan, Casimir densities induced
by a sphere in the hyperbolic vacuum of de Sitter spacetime, Phys. Rev. D 
\textbf{104}, 065017 (2021).

\bibitem{Henr55} P. Henrici, Addition theorems for general Legendre and
Gegenbauer tunctions, J. Ration. Mech. Anal. \textbf{4}, 983 (1955).

\bibitem{Bles09} A.C. Bleszynski-Jayich, W.E. Shanks, B. Peaudecerf, E.
Ginossar, F. von Oppen, L. Glazman, J.G.E. Harris, Persistent currents in
normal metal rings, Science \textbf{326}, 272 (2009).

\bibitem{Bluh09} H. Bluhm, N.C. Koshnick, J.A. Bert, M.E. Huber, M.A. Moler,
Persistent currents in normal metal rings, Phys. Rev. Lett. \textbf{10}2,
136802 (2009).

\bibitem{Cast13} M.A. Castellanos-Beltran, D.Q. Ngo, W. E. Shanks, A.B.
Jayich, J.G.E. Harris, Measurement of the full distribution of persistent
current in normal-metal rings, Phys. Rev. Lett. \textbf{110}, 156801 (2013).

\bibitem{Sark25} S. Sarkar, S. Satpathi, S.K. Pati, Enhancement of
persistent current in a non-Hermitian disordered ring, arXiv:2502.12805.

\bibitem{Duns24} T.M. Dunster, Simplified uniform asymptotic expansions for
associated Legendre and conical functions, arXiv:2410.03002.

\bibitem{Prud2} A.P. Prudnikov, Yu.A. Brychkov, O.I. Marichev, \textit{%
Integrals and Series} (Gordon and Breach, New York, 1986), Vol. 2.

\bibitem{Wats66} G.N. Watson, \textit{A Treatise on the Theory of Bessel
Functions} (Cambridge University Press, Cambridge, 1966).

\bibitem{Pfau82} J. D. Pfautsch, A new vacuum state in de Sitter space,
Phys. Lett. B \textbf{117}, 283 (1982).

\bibitem{Sasa95} M. Sasaki, T. Tanaka, K. Yamamoto, Euclidean vacuum mode
functions for a scalar field on open de Sitter space, Phys. Rev. D \textbf{51%
}, 2979 (1995).
\end{thebibliography}
\end{document}